\documentclass[aps,pra,floats,twocolumn]{revtex4}

\usepackage[latin1]{inputenc}
\usepackage{graphicx}

\usepackage{amsmath,amsthm,amsfonts,eufrak,amssymb}
\input amssym.def
\newsymbol\square 1003

\newcommand{\jd}{\tfrac{1}{2}}

\newcommand{\ket}[1]{\ensuremath{|#1\rangle}}

\begin{document}

\title{Generalized spin squeezing inequalities in $N$-qubit systems: theory and experiment.}
\author{J. K. Korbicz$^{1,2}$, O. G\"uhne$^{3}$, M. Lewenstein$^{2,1}$, H. H\"affner$^{4,3}$, C. F. Roos$^{4,3}$, and R. Blatt$^{4,3}$}
\affiliation{$^1$ Institut f\"ur Theoretische Physik, Universit\"at Hannover, Appelstra\ss e 2, D-30167
Hannover, Germany}

\affiliation{$^2$ ICREA and ICFO -- Institut de Ci\`{e}ncies Fot\`{o}niques, Mediterranean Technology Park, 08860
Castelldefels (Barcelona), Spain}
 
\affiliation{$^3$ Institut f\"ur Quantenoptik und Quanteninformation,
\"Osterreichische Akademie der Wissenschaften, A-6020 Innsbruck,
Austria}

\affiliation{$^4$ Institut f\"ur Experimentalphysik, Universit\"at Innsbruck,
Technikerstra\ss e 25, A-6020 Innsbruck,
Austria}

\begin{abstract}
We present detailed derivations, various improvements and application to 
concrete experimental data of spin squeezing inequalities 
formulated recently by some of us [Phys. Rev. Lett. {\bf 95}, 120502 (2005)]. These 
inequalities  generalize the concept
of the spin squeezing parameter,  and provide necessary and
sufficient conditions for genuine 2-, or 3- qubit entanglement for
symmetric states, and sufficient condition for general $N$-qubit
states. We apply our method to theoretical study of Dicke states, and, in particular, 
to $W$-states of $N$ qubits. Then, we analyze the recently
experimentally generated $7$- and $8$-ion $W$-states [Nature  {\bf 438}, 643 (2005)].
We also present some novel details concerning this experiment. 
Finally, we improve criteria for detection of genuine tripartite entanglement
based on entanglement witnesses. 
\end{abstract}

\maketitle

\section{Introduction}

Experimental generation and characterization of entanglement on a macroscopic, 
or mesoscopic scales seem to be one of the necessary prerequisites of scalable 
quantum information processing. A spectacular progress 
has been achieved   recently in the
area of quantum correlated systems of atoms, and in particular {\it
macroscopic atomic ensembles} \cite{hald}. The main goal of these studies is to achieve an efficient quantum 
interface between light 
and atoms with spin, or pseudo-spin internal states, using the generalized quantum Faraday 
effect.
Such settings already 
allowed one to demonstrate entanglement of distant 
atomic objects \cite{jul}, or deterministic memory for light \cite{jul2} that 
can be retrieved using quantum teleportation \cite{tele}.  
Entanglement between light and atoms, and between atoms themselves plays, of course, 
essential role in these experiments. 

It worth stressing that the light-atoms interface based on using the quantum Faraday effect 
does not only allow one to 
measure  and detect atomic states.
It does also provide a tool for manipulations and engineering of quantum fluctuations of atomic 
spins. The latter possibility might be 
of fundamental importance for the future implementations of distributed
 quantum information processing. In particular, the methods 
of atomic ensembles can be carried over to another rapidly developing 
area of ultracold atomic gases. Here, the interest would be to measure
characterize, and finally engineer quantum fluctuations of the total 
atomic spin in spinor ultracold gases (for a review see \cite{review}) 
that has been 
intensively studies since the seminal theory papers of T.-L. Ho 
\cite{Ho98} and T. Ohmi and K. Machida \cite{machida},  
as well as the experiments 
performed by the MIT group on optically trapped sodium 
Bose-Einstein condensates (BEC) 
\cite{Sten98}. Particularly interesting are prospects of applications 
of these methods to strongly correlated states of spin ultracold gases 
in optical lattices \cite{demler}.

Yet another rapidly developing related area is that of quantum information 
processing with trapped ions. 
After the first
works, in which the 3- and 4-ion $GHZ$-state \cite{wineland4},  
and 3-ion $W$- and $GHZ$-state \cite{blatt3} 
have been generated \cite{photons}, in recent 
experiments the  tomography of 6-, 7-, and 8-ion $W$-states has 
been performed \cite{blatt8}, and the 6-ion $GHZ$-state has been 
generated \cite{wineland6}. 

The problem of characterization of the generated forms of multipartite 
entanglement \cite{ent}, or more generally, 
of characterization of many-body quantum correlations 
is thus of essential importance for the investigations of such mesoscopic  systems. 
One of the possible ways to achieve it, is to measure the total spin
(or pseudo-spin) of  atoms (or ions) and its quantum fluctuations, which 
is, of course,  possible by performing state tomography. The central role 
in this approach, applied to atomic ensembles, has been played so far
by the, so called, spin squeezing parameter $\xi^2$, 
introduced by M. Kitagawa and M. Ueda in Ref. \cite{ssparam1}.
As it was shown in Refs. \cite{ssparam2,sanders} it provides
a sufficient entanglement criterion for atomic ensembles. 
On top of that, $\xi^2$ is particularly appreciated by experimentalists 
since: i) it has a clear physical meaning,
ii) it can be relatively easy measured, iii) it
is defined by a simple operational expression, iv) 
it provides a figure of merit for atomic clocks.  
However, until our recent Letter \cite{naszprl} 
no further investigations  to relate $\xi^2$ to other concepts
of quantum information have been carried out.

The present work is a substantially extended version
of Ref. \cite{naszprl}. Apart from the expanded 
theoretical analysis, we present here a  
detailed description of the ion-trap experiment of Ref. \cite{blatt8},
to which output we apply our inequalities. 

Let us first recall that in Ref. \cite{naszprl} we  have generalized and connected 
the concept of
spin squeezing parameters to the theory of entanglement witnesses
\cite{witnesses}, i. e. such observables $\mathcal{W}$ that have
non-negative averages for all separable states and there exists
an entangled state $\varrho$ such that
$\text{tr}\big(\varrho\mathcal{W}\big)<0$.
In order to derive the generalized spin squeezing inequalities,
we have proposed  a general method of expressing  state
averages of the appropriate
entanglement witnesses in terms of the macroscopic
spin operators:
\begin{equation}\label{S}
J^i=\sum_{a=1}^N \frac{1}{2}\sigma^i_{a}\,\,, \ \ i=1,2,3 
\end{equation}
(we work in the units $\hbar=1$). 
Here by $\sigma^i$ we denote Pauli matrices, indices $a,b,c\dots$
enumerate the particles of the ensemble, and 
$\sigma^i_{a}=\sigma^i\otimes {\bf 1}_{1\dots\hat{a}\dots N}$ 
(hat over the
index denotes that it is omitted). It is worth recalling
at this place that in the standard terminology \cite{ssparam1} 
a state of a spin-$J$ system is called spin
squeezed if there exists a direction ${\bf n}$, orthogonal to the
mean spin $\langle {\bf J}\rangle$, such that:
\begin{equation}
\xi^2=2\langle \Delta J_{\bf n}^2\rangle/J <1,\label{xi}
\end{equation}
where $J_{\bf n}={\bf n \cdot J}$.

Our method works as follows: we begin with considering symmetric
states of $N$ qubits first, i.e. states $\varrho$ that satisfy: 
\begin{equation}\label{symetr}
P
\varrho P = \varrho,
\end{equation}
where $P$ is an orthogonal projector onto the
symmetrized product of individual qubit spaces
$\mathcal{H}_s=\text{Sym}(\mathbb{C}^2\otimes\dots\otimes\mathbb{C}^2)$
($\text{Sym}$ denotes symmetrization). It is known that for symmetric states of two and three qubits
the necessary and sufficient condition for  separability of a quantum state is equivalent
to the positivity of partial transpose (PPT) of the state.
For two qubits the PPT condition is in fact the necessary and sufficient condition for separability of 
arbitrary (also non-symmetric) states \cite{PPT}; for symmetric states 
of three qubits this result  has been shown in Ref. 
\cite{eckert}. The knowledge of the necessary and sufficient separability 
criterion allows us to  derive the
complete families of generalized spin squeezing inequalities,
which provide necessary and sufficient conditions for genuine 2-,
or 3- qubit entanglement for symmetric states.

Our inequalities at the same time
provide a sufficient condition for entanglement of general,
i.e. not necessarily symmetric, states of $N$
qubits \cite{notka}. The  results of Ref. \cite{naszprl} 
imply also that, if we somewhat broaden the standard notion 
of spin squeezing (\ref{xi}), then for spin-$J$ systems 
represented as a collection of $2J$ qubits,
 spin squeezing becomes equivalent to the bipartite entanglement 
among the qubits (see also \cite{sanders} where the 
implication in one direction was obtained). We also derive and discuss  
 improved w.r.t. Ref. \cite{naszprl} versions of somewhat simpler 
spin squeezing inequalities that provide sufficient conditions 
for genuine 3-qubit  entanglement.

To prepare the necessary data for the analysis of the output of the experiment 
from Ref. \cite{blatt8}, as well as to show how to obtain our inequalities for
concrete purposes, we present in this paper a very explicit derivation of the 
inequalities for the, so called, 
Dicke  states \cite{Dicke}, sometimes also called generalized $W$-states. 
We show step-by-step how to derive the inequalities probing genuine 
$2$- and $3$-qubit entanglement of this states. We also calculate all the necessary data data for checking $7$- and $8$-qubit $W$ states, which are of particular interest for us. 

In the part of our work dedicated to the experiment, we present a detailed description and
analysis of the experimental production and state tomography of 6-, 7-, and 8-particle $W$-states of trapped ions, first reported in Ref. \cite{blatt8}. Here, we describe the details of the production of the states in an ion trap, dedicated to quantum information processing \cite{QC-HowTo}. We explain the step-by-step generation algorithm, that was implemented in the experiment. We then explain how the state tomography was performed, and show the full reconstructed density matrix of the $7$-qubit $W$-state. We analyze the experimental imperfections as well. Finally, we apply our spin squeezing inequalities to the experimental data to confirm the presence of 2- and 3-qubit entanglement in the generated states.

Let us stress that all of the proposed novel inequalities,
analogously  as the previously known squeezing parameter, i) have a clear physical
meaning in terms of generalized squeezing and entanglement
conditions, ii) can be relatively easy measured, and iii) are
given by complex, but {\it elementary}  expressions. Although in this paper we apply our theoretical  tools to a fully restored density matrix from Ref. \cite{blatt8}, it is very important to understand that these tools require measurements of low order moments of the total spin fluctuations only. Hence, checking of our inequalities can be relatively directly performed in large systems, such as atomic ensembles, where, in general, quantum tomography is not feasible.

We also note that recently G.~T\'oth has also derived various 
types of entanglement criteria based on entanglement witnesses 
and on the uncertainty of collective observables, such 
as the total spin or energy \cite{toth}. These criteria are
useful to detect the, so-called, cluster states and many body 
singlet states, but they may also be used to detect the Dicke 
states discussed in this paper. 

The work is organized as follows: in Sections II and III we 
revise the derivation of 2- and 3-qubit entanglement criteria.
 We give there some calculational details
as well as correct versions of the formulas (17), (18) and (19) 
from  Ref. \cite{naszprl}. These results are very general and apply to any system of qubits in any state: 
from few ions, through atomic ensembles to ultracold spinor gases.  
In Section IV we specify our inequalities to a concrete example
of experimentally accessible Dicke states,  and show how to construct our criteria in this case. 
In particular, we provide here explicit 
data for the case of 7- and 8-qubit $W$-states, preparing the input data for the analysis of the experiment of Ref. \cite{blatt8}. Section V  
is devoted to  a detailed description of this experiment. 
We apply here our inequalities to the analysis of the  
output of this experiment, confirming the presence of 2- and 3-qubit entanglement.  
Section VI is dedicated to the construction of
simplified witnesses detecting genuine 3-qubit entanglement, improving 
the similar witnesses we 
constructed earlier in Ref. \cite{naszprl}. The simplified witnesses, unfortunately, are not very useful for 
$W$-states, since they detect entanglement only for low number  of qubits. That is why we do not use them for analysis of the 
experimental data of Ref. \cite{blatt8}. 
We summarize our results in Section VII.

\section{Detection of bipartite entanglement}

In this Section we present a detailed derivation of generalized squeezing inequalities that detect 2-qubit entanglement. 
The aim is to use the quantum fluctuations of the total spin, whose low moments can be relatively easily measured, as an indicator of entanglement. The results obtained in this Section, as well as the method itself,  have a
 very general character, and can be used for arbitrary systems of qubits in any quantum state.  

Let us recall that a multiqubit state $\varrho$ possesses 2-qubit 
entanglement if for some
qubits $a$ and $b$ the reduced density matrix:
\begin{equation}
\varrho_{ab}=\text{tr}_{1..\hat{a}..\hat{b}..N}\varrho
\end{equation}
is entangled. The PPT criterion \cite{PPT} implies that $\varrho_{ab}$ is
entangled iff there exists a vector $|\psi\rangle$ such that:
\begin{equation}\label{PPT}
\text{tr}_{ab}\big(\varrho_{ab}|\psi\rangle\langle\psi|^{T_1}\big)<0,
\end{equation}
where transpose is defined with respect to (w.r.t.) the standard basis $|0\rangle,|1\rangle$. 
As $\psi$ we can take any eigenvector of $\varrho_{ab}^{T_1}$ corresponding to a negative eigenvalue.

According to our general strategy we first consider symmetric states, as then we 
can obtain a convenient parametrization of $|\psi\rangle$. In the 2-qubit case 
we can take advantage of the low dimensionality and use the explicit form of 
$\varrho_{ab}^{T_1}$. Let us first fix the basis of each qubit space by 
$\sigma^z|0\rangle=|0\rangle$, $\sigma^z|1\rangle=-|1\rangle$. Then we have 
that:
\begin{equation}
\varrho^{T_1}=\left [ \begin{array}{cccc} \epsilon_0 & \delta & \delta^* & \tau\\
                                      \delta^* & \epsilon_1 & \varpi^* & \varsigma^* \\
                      \delta & \varpi & \epsilon_1 & \varsigma\\
                      \tau & \varsigma & \varsigma^* & \epsilon_2 \end{array}\right],
\end{equation}
where $\epsilon_0,\epsilon_1,\epsilon_2, \tau \in \mathbb{R}$. It
is easy to check that vectors of the type: 
\begin{equation}\label{type}
|\psi\rangle= \eta |00\rangle+\beta |01\rangle+\beta^*|10\rangle
+\gamma |11\rangle,\quad \eta,\gamma\in\mathbb{R} 
\end{equation}
are preserved by
$\varrho^{T_1}$, and, since they have three independent parameters
(we take them to be normalized, although it is not important for the condition (\ref{PPT})),
it is possible to find a solution
of the eigenvalue equation. Hence, the negative eigenvalue vector in the inequality (\ref{PPT})
must be of this form. From Eq. (\ref{type}) it follows that the matrix $[\psi]$ of coefficients
of $|\psi\rangle$ is hermitian:  
\begin{equation}\label{form}
[\psi]=\left [ \begin{array}{cc} \eta & \beta\\
                                 \beta^* & \gamma \end{array}\right],
\end{equation}
and hence we can diagonalize it by some $\tilde{U}\in SU(2)$ (modulo $U(1)$ phase rotation):
\begin{equation}
[\psi]=\tilde{U}^\dagger\Delta\tilde{U}. 
\end{equation}
Note that due to the normalization of $\ket{\psi}$, the eigenvalue matrix $\Delta$ can be put in the following form:
\begin{equation}\label{eigenval}
\Delta=\left [ \begin{array}{cc}\text{sin}\frac{\varphi}{2} & 0 \\
                                 0 & \pm\text{cos}\frac{\varphi}{2} \end{array}\right],
\quad -\pi\le \varphi \le \pi.
\end{equation}
Rewriting Eq. (\ref{form}) explicitly in the basis, and using Eq. (\ref{eigenval}), 
we finally obtain the following parametrization from Ref. \cite{naszprl}:
\begin{equation}\label{psi2}
|\psi\rangle=U^*\otimes U |\psi_0\rangle\,,\ \ |\psi_0\rangle=
\text{sin}\frac{\varphi}{2} |00\rangle+\text{cos}\frac{\varphi}{2}|11\rangle\,,
\end{equation}
where $U=\tilde{U}^T$, and we have fixed the overall phase. The parameters 
$\eta,\beta,\gamma$ from the decomposition (\ref{type}) 
are now encoded into $\varphi$ and $U$. Using the above parametrization 
inequality (\ref{PPT}) takes the following form:
\begin{equation}\label{PPT2}
\text{tr}_{ab}\big(\varrho_{ab}U \otimes 
U |\psi_0\rangle\langle\psi_0|^{T_1}U^\dagger \otimes U^\dagger\big)<0\,.
\end{equation}

In order to rewrite the condition (\ref{PPT2}) with the total spin operators 
(\ref{S}), we first recall that $|\psi_0\rangle\langle\psi_0|^{T_1}$ 
can be decomposed into Pauli matrices, as it was done in Ref. \cite{naszprl}.
Then, the adjoint action of $SU(2)$ in the inequality (\ref{PPT2}) induces 
a $SO(3)$ rotation $R$ of the Pauli matrices: 
$U \sigma^i U^\dagger=R^i_{\phantom{i}j}\sigma^j$ (here and 
throughout the work we sum over the repeated indices). We will denote the 
axes of the rotated frame by $\bf{k},\bf{l},\bf{n}$.

Since in the symmetric case we currently consider all the reductions 
$\varrho_{ab}$ are of the same form,
we can sum the inequalities (\ref{PPT2}) over all pairs of qubits: $\sum_{\langle
ab\rangle}=\sum_{a=1}^{N-1}\sum_{b=a+1}^{N}$, without affecting the inequality sign. 
However, before we do so, we extend $|\psi\rangle\langle\psi|^{T_1}$ 
from the space of the qubits $ab$ to the full Hilbert space of $N$ qubits by:
$|\psi\rangle_{ab}\langle\psi|^{T_1}=|\psi\rangle\langle\psi|^{T_1}\otimes{\bf 1}_{1..\hat{a}..\hat{b}..N}$. 
Then, we obtain that:
\begin{equation}\label{gowno}
\sum_{\langle ab \rangle}\text{tr}_{ab}
\big(\varrho_{ab}|\psi\rangle\langle\psi|^{T_1}\big)=\text{tr}\big(\varrho
\sum_{\langle ab \rangle}|\psi\rangle_{ab}\langle\psi|^{T_1}\big). 
\end{equation}
Now we can plug the Pauli matrix decomposition of $|\psi\rangle\langle\psi|^{T_1}$ 
into Eq. (\ref{gowno}), and, using the identity:
\begin{equation}
\sum_{\langle ab\rangle}\sigma ^i_a\otimes\sigma ^i_b=2(J^i)^2-\frac{N}{2},
\end{equation}
obtain the desired form of the condition (\ref{PPT}), i.e. a symmetric state $\varrho$ possesses 
bipartite entanglement iff there exist $-\pi\le \varphi \le \pi$ and $U\in SU(2)/U(1)$, such that 
the following inequality holds:
\begin{eqnarray}
& & \text{sin}\varphi\Big[\langle J^2_{\bf k}\rangle+\langle
J^2_{\bf l}\rangle-\frac{N}{2}\Big]-(N-1)\text{cos}\varphi \langle J_{\bf n} \rangle\nonumber\\
& &+\langle J^2_{\bf n}\rangle +\frac{N(N-2)}{4}<0,\label{2x2}
\end{eqnarray}
where all the averages are taken w.r.t. the full state $\varrho$.

In case of a general, i.e. not necessarily symmetric, state $\varrho$ 
observe that, if there exist $-\pi\le \varphi \le \pi$ and $U\in SU(2)/U(1)$
the same for all pairs of qubits, and such that the sum (\ref{gowno}) is negative, 
then there must be at least one pair $ab$ for which 
$\text{tr}_{ab}\big(\varrho_{ab}|\psi\rangle\langle\psi|^{T_1}\big)<0$, 
and hence the state $\varrho$ possesses bipartite entanglement.
Thus, the condition (\ref{2x2}) is also a sufficient condition for bipartite entanglement
for general states.
 
For a given negative eigenvalue vector $|\psi\rangle$ the left hand side 
of the inequality (\ref{2x2}) is completely determined.
However, we can also treat it as a function of the parameters of $|\psi\rangle$, 
and as such it can be optimized. 
In particular, keeping the frame ${\bf k,l,n}$ fixed, we can search for 
the minimum w.r.t. $\varphi$. 
Let us call this minimum $\varphi_0$. Clearly, if the inequality (\ref{2x2}) is satisfied 
for some $\varphi$, then it will   
be also satisfied for $\varphi_0$, and vice versa. Hence, it is enough to 
check the condition (\ref{2x2}) only for $\varphi_0$.
Performing the minimization, we obtain that:
\begin{equation}
\text{sin}\varphi_0 = -\frac{\langle J^2_{\bf k}\rangle+\langle
J^2_{\bf l}\rangle-\frac{N}{2}}{\sqrt{\big[\langle J^2_{\bf k}\rangle+\langle
J^2_{\bf l}\rangle-\frac{N}{2}\big]^2+(N-1)^2\langle J_{\bf n}\rangle^2}}
\end{equation}
\begin{equation}
\text{cos}\varphi_0 = \frac{(N-1)\langle J_{\bf n}\rangle}{\sqrt{\big[\langle J^2_{\bf k}\rangle+\langle
J^2_{\bf l}\rangle-\frac{N}{2}\big]^2+(N-1)^2\langle J_{\bf n}\rangle^2}},
\end{equation}
and the inequality (\ref{2x2}) becomes:
\begin{equation}\label{2x2min}
\langle J^2_{\bf n}\rangle +\frac{N(N-2)}{4}<\sqrt{\Big[\langle
J^2_{\bf k}\rangle+\langle J^2_{\bf
l}\rangle-\frac{N}{2}\Big]^2+(N-1)^2\langle J_{\bf n}
\rangle^2}\,.
\end{equation}
As a result, we arrive at the \cite{naszprl}:

{\bf Criterion for bipartite entanglement.} {\it If there exist
mutually orthogonal directions
 $\bf{k}$, $\bf{l}$, $\bf{n}$ such that the inequality (\ref{2x2min}) 
holds, then the state $\varrho$
 possesses bipartite entanglement. For symmetric states the above 
condition is both necessary and sufficient.}

In the latter case, due to the equality:
\begin{equation}
\langle J_{\bf k}^2\rangle +\langle J_{\bf l}^2\rangle +\langle J_{\bf n}^2\rangle = \frac{N(N+2)}{4},
\end{equation}
the criterion (\ref{2x2min}) can be simplified to:
\begin{equation}\label{2x2'}
\frac {4 \langle \Delta J_{\bf n}^2\rangle}{N}< 1- \frac{4\langle
J_{\bf n} \rangle^2}{N^2}.
\end{equation}

The relation of the criterion (\ref{2x2min}) to the standard spin squeezing
condition (\ref{xi}) is the following. Spin-$J$ state can be equivalently
represented as a symmetric state of $N=2J$ qubits. Intuitively, spin squeezing 
should refer to the existence of non-classical correlations among the
qubits \cite{ssparam1}. Indeed, the criterion (\ref{2x2'}) provides a rigorous
proof for this intuitive picture, as, on one hand, if the  
condition (\ref{xi}) is satisfied, then the inequality (\ref{2x2'})
is satisfied as well, since in this particular case 
$\langle J_{\bf n} \rangle=0$ and $J= N/2$. Hence,
spin-$J$ squeezed states possess 2-qubit entanglement
\cite{sanders}. On the other hand, if we broaden the standard definition of spin squeezing
 (\ref{xi}), and allow 
the direction ${\bf n}$ to be arbitrary, then we also obtain the converse statement:
the condition (\ref{2x2'}) implies the existence of a spin component $J_{\bf n}$,
such that $\langle \Delta J_{\bf n}^2\rangle<J/2$. Note however, that from the condition
(\ref{2x2'}) it does not follow that the direction of squeezing ${\bf n}$ is orthogonal 
to $\langle {\bf J} \rangle$. Thus, we obtain a more general type of squeezing. 
In Section IV we will show somewhat extreme examples
of state, for which ${\bf n}$ is actually parallel to the mean spin.

\section{Detection of tripartite entanglement}
In the previous Section we have shown that the presence of bipartite entanglement is detected by the second order moments of 
the total spin. Given our methods, it is natural to expect that the tripartite entanglement should be detectable by  
third order moments of ${\bf J}$. In this Sections we derive the corresponding generalized squeezing inequalities in the most 
generic form, valid for arbitrary quantum states of systems of qubits. 

As in the previous Section, we begin with considering symmetric states first. Recall that the PPT criterion 
still works for the tripartite reductions $\varrho_{abc}$ of such states \cite{eckert}, and there are two 
families of potential negative-eigenvalue-vectors of $\varrho_{abc}^{T_1}$ \cite{Dur, naszprl}:
\begin{eqnarray}
& & |\psi\rangle = A \otimes B \otimes B |GHZ_3\rangle,  \label{GHZ}\\
& & |\psi\rangle = A \otimes U \otimes U |W_3\rangle \label{W} \,.
\end{eqnarray}
Here, matrices $A,B \in SL(2,\mathbb{C})$, $U\in SU(2)$, and
$|GHZ_3\rangle=(1/\sqrt{2})(|000\rangle+|111\rangle)$,
$|W_3\rangle=(1/\sqrt{3})(|001\rangle+|010\rangle+|100\rangle)$. The
action of $SL(2,\mathbb{C})$ on the Pauli matrices in the
decomposition of $|\psi\rangle\langle\psi|^{T_1}$ now induces
restricted, i.e. orientation and time-orientation preserving,
Lorentz transformations:
\begin{equation}\label{Lorentz}
A^*\sigma^\mu A^T=\Lambda^\mu_{\phantom{\mu}\nu}\sigma^\nu\,, \ B\sigma^\mu B^\dagger=L^\mu_{\phantom{\mu}\nu}\sigma^\nu\,,\ \sigma^0={\bf 1}\,
\end{equation}
(Greek indices run through $0\dots 3$), and the PPT condition takes the following form:
\begin{equation}\label{poly2}
\text{tr}_{abc}\big(\varrho_{abc}|\psi\rangle\langle\psi|^{T_1}\big)=\tfrac {1}{8} K_{\alpha\beta\gamma}\text{tr}_{abc}\big(\varrho_{abc} \sigma^{\alpha} \otimes \sigma^\beta \otimes \sigma^{\gamma}\big) < 0 
\end{equation}
(note the summation convention). Tensor $K_{\alpha\beta\gamma}$ reads:
\begin{eqnarray}
& & K_{\alpha\beta\gamma}(\Lambda,L)\!= \!\Lambda^0_{\phantom{\mu}\alpha} L^0_{\phantom{\mu}\beta}L^0_{\phantom{\mu}\gamma}\!\!+\!\Lambda^0_{\phantom{\mu}\alpha} L^3_{\phantom{\mu}\beta}L^3_{\phantom{\mu}\gamma}\!\!+\!\Lambda^1_{\phantom{\mu}\alpha} L^1_{\phantom{\mu}\beta} L^1_{\phantom{\mu}\gamma}\nonumber\\
& & + 2 \Lambda^3_{\phantom{\mu}\alpha} L^0_{\phantom{\mu}(\beta} L^3_{\phantom{\mu}\gamma)}-\Lambda^1_{\phantom{\mu}\alpha} L^2_{\phantom{\mu}\beta} L^2_{\phantom{\mu}\gamma}+ 2\Lambda^2_{\phantom{\mu}\alpha} L^1_{\phantom{\mu}(\beta} L^2_{\phantom{\mu}\gamma)}\,\label{KGHZ}
\end{eqnarray}
for the $GHZ$-family (\ref{GHZ}), and \cite{corr1}:
\begin{eqnarray}
& & K_{\alpha\beta\gamma}(\Lambda,R)= \tfrac{1}{3}\Big\{3\Lambda^0_{\phantom{\mu}\alpha} R^0_{\phantom{\mu}\beta}R^0_{\phantom{\mu}\gamma}-\!3\Lambda^3_{\phantom{\mu}\alpha} R^3_{\phantom{\mu}\beta}R^3_{\phantom{\mu}\gamma}\nonumber\\
& & \!+ 2\Lambda^0_{\phantom{\mu}\alpha} R^0_{\phantom{\mu}(\beta}R^3_{\phantom{\mu}\gamma)}\!+\!\Lambda^3_{\phantom{\mu}\alpha} R^0_{\phantom{\mu}\beta}R^0_{\phantom{\mu}\gamma}\!-\!\Lambda^0_{\phantom{\mu}\alpha} R^3_{\phantom{\mu}\beta}R^3_{\phantom{\mu}\gamma}\label{KW}\\
& & \!-2\Lambda^3_{\phantom{\mu}\alpha} R^0_{\phantom{\mu}(\beta}R^3_{\phantom{\mu}\gamma)}\!+\!4\Lambda^1_{\phantom{\mu}\alpha} R^0_{\phantom{\mu}(\beta}R^1_{\phantom{\mu}\gamma)}\!+\!4\Lambda^1_{\phantom{\mu}\alpha} R^1_{\phantom{\mu}(\beta}R^3_{\phantom{\mu}\gamma)}\nonumber\\
& &
-4\Lambda^2_{\phantom{\mu}\alpha}R^0_{\phantom{\mu}(\beta}R^2_{\phantom{\mu}\gamma)}-4\Lambda^2_{\phantom{\mu}\alpha}R^2_{\phantom{\mu}(\beta}R^3_{\phantom{\mu}\gamma)}+2\Lambda^0_{\phantom{\mu}\alpha}R^1_{\phantom{\mu}\beta}R^1_{\phantom{\mu}\gamma}\nonumber\\
& &
+2\Lambda^3_{\phantom{\mu}\alpha}R^1_{\phantom{\mu}\beta}R^1_{\phantom{\mu}\gamma}+2\Lambda^3_{\phantom{\mu}\alpha}R^2_{\phantom{\mu}\beta}
R^2_{\phantom{\mu}\gamma}+2\Lambda^0_{\phantom{\mu}\alpha}
R^2_{\phantom{\mu}\beta}R^2_{\phantom{\mu}\gamma}\Big\}\,\nonumber
\end{eqnarray}
for the $W$-family (\ref{W}). Here, $R^{\mu}_{\phantom{\mu}\nu}$ is the four-dimensional 
embedding of the rotation generated by $U$ from Eq. (\ref{W}), and the round brackets $(\ )$ 
around indices denote symmetrization, e.g. $A^{(ij)} = (A^{ij} + A^{ji})/2.$ 
Note that the relativistic notation is used only for our convenience. 
We could have as well put all the indices at the same level as we are {\emph not} 
going to lower or rise them with the Minkowski metric. 

Next, we sum the inequalities (\ref{poly2}) over all triples of qubits:
$\sum_{\langle abc\rangle}=
\sum_{a=1}^{N-2}\sum_{b=a+1}^{N-1}\sum_{c=b+1}^N$, just like we summed the
inequalities (\ref{PPT2}) in the previous Section:
\begin{eqnarray}
& &\sum_{\langle abc\rangle}K_{\alpha\beta\gamma}\text{tr}_{abc}\big(\varrho_{abc} \sigma^{\alpha} \otimes \sigma^\beta \otimes \sigma^{\gamma}\big)=\nonumber\\
& &K_{\alpha\beta\gamma}\text{tr}\big(\varrho\sum_{\langle abc\rangle}\sigma_a^{\alpha} \otimes \sigma_b^\beta \otimes \sigma_c^{\gamma}\big).
\label{suma3}\end{eqnarray}
Because of the symmetry condition (\ref{symetr}), we can rewrite Eq. (\ref{suma3}) as follows:
\begin{equation}\label{suma3'}
K_{\alpha\beta\gamma}\text{tr}\big(\varrho P\sum_{\langle abc\rangle}\sigma_a^{\alpha} \otimes \sigma_b^\beta \otimes \sigma_c^{\gamma} P^\dagger\big),
\end{equation}
and observe that due to the action of $P$, we can substitute $\sigma_a^{\alpha} \otimes \sigma_b^\beta \otimes \sigma_c^{\gamma}$ with the symmetrized product $\sigma_a^{(\alpha} \otimes \sigma_b^\beta \otimes \sigma_c^{\gamma)}$. This finally allows us to rewrite
Eq. (\ref{suma3'}) with the total spin operators $J^i$ (supplemented by an artificial ``time-component'' 
$J^0=(N/2){\bf 1}$ for compactness of the notation), because of the identity:
\begin{eqnarray}
& & 3 \sum_{\langle abc\rangle} \sigma_a^{(\alpha}\otimes\sigma_b^\beta\otimes\sigma_c^{\gamma)}=
4 J^{(\alpha}J^\beta J^{\gamma)}-6 f^{(\alpha\beta}_{\phantom{(\alpha\beta}\mu} J^{(\gamma)}J^{\mu)}\nonumber\\
& &
+2f^{(\alpha\beta}_{\phantom{(\alpha\beta}\mu}f^{(\gamma)\mu)}_{\phantom{(\alpha\beta}\nu}J^\nu-
f^{(\alpha\beta}_{\phantom{(\alpha\beta}\mu}f^{[\gamma)\mu]}_{\phantom{(\alpha\beta}\nu}J^\nu\,.\label{masakra}
\end{eqnarray}
The symmetrization above is taken w.r.t. $\alpha\beta\gamma$ and $\gamma\mu$ separately and 
square brackets $[\ ]$ around Greek indices denote antisymmetrization, e.g. $A^{[ij]} = (A^{ij} - A^{ji})/2$. The constants 
$f^{\alpha\beta}_{\phantom{0\alpha}\gamma}$
are defined through: $\sigma^\mu\sigma^\nu=f^{\mu\nu}_{\phantom{0\alpha}\gamma}\sigma^\gamma$. 
Their numerical values are as follows: $f^{0\alpha}_{\phantom{\alpha}\beta}=f^{\alpha
0}_{\phantom{0\alpha}\beta}=\delta^\alpha_{\phantom{\alpha}\beta}$,
$f^{ij}_{\phantom{jk}\alpha}=\text{i}\sum_l\epsilon^{ijl}\delta^l_{\phantom{l}\alpha}+\delta^{ij}\delta^0_{\phantom{0}\alpha}$. Substituting Eq. (\ref{masakra}) into Eq. (\ref{suma3'}) leads us to the 
\cite{corr2}:

{\bf Criterion for tripartite entanglement.} {\it A symmetric state $\varrho$ possesses 
genuine tripartite entanglement iff there exist two restricted Lorentz 
transformations 
$\Lambda$, $L$, or a restricted Lorentz transformation $\Lambda$ and a 
rotation $R$, such that:
\begin{eqnarray}
X(\varrho) &\equiv& K_{(\alpha\beta\gamma)}\Big\{2\langle J^{\alpha}J^\beta
J^{\gamma}\rangle
-3f^{\alpha\beta}_{\phantom{\alpha\beta}\mu}\langle
J^{(\gamma}J^{\mu)}\rangle\label{2x2x2} \\
&+&f^{\alpha\beta}_{\phantom{\alpha\beta}\mu}f^{(\gamma\mu)}_{\phantom{(\alpha\beta)}\nu}\langle
J^\nu\rangle\nonumber-\frac{1}{2}f^{\alpha\beta}_{\phantom{\alpha\beta}\mu}f^{[\gamma\mu]}_{\phantom{(\alpha\beta}\nu}\langle
J^\nu \rangle \Big\}<0
\end{eqnarray}
holds, with $K_{\alpha\beta\gamma}$ given by Eq. (\ref{KGHZ}), or by Eq. (\ref{KW}).}

For a general state $\varrho$ we could, as in the previous Section, generate a sufficient 
entanglement condition by applying the same witness $|\psi\rangle\langle\psi|^{T_1}$, 
with $|\psi\rangle$ given by Eq. (\ref{GHZ}) or by Eq. (\ref{W}), to all tripartite reductions 
$\varrho_{abc}$. However, then we cannot use the symmetry arguments like we used in 
Eq. (\ref{suma3'}), and directly apply the identity (\ref{masakra}). Instead, we construct 
from the families (\ref{GHZ}) and (\ref{W}) different witnesses, given for the $W$-family (\ref{W}) by: 
{\setlength \arraycolsep{0pt}
\begin{eqnarray}
\frac{1}{3}\Big\{&\big(&AUU|W_3\rangle\langle W_3|A^\dagger U^\dagger U^\dagger\big)^{T_1}\nonumber\\
&+&\big(UAU|W_3\rangle\langle W_3|U^\dagger A^\dagger U^\dagger\big)^{T_2}\nonumber\\
&+& \big(UUA|W_3\rangle\langle W_3|U^\dagger U^\dagger A^\dagger \big)^{T_3} \Big\}\label{skurwisyn},
\end{eqnarray}}
(we have omitted tensor product signs, $\otimes$, for compactness), and analogously 
for the $GHZ$-family (\ref{GHZ}). We then apply the witnesses 
(\ref{skurwisyn}) to all tripartite reductions of $\varrho$, which 
effectively leads to the substitution of $K_{\alpha\beta\gamma}$ by $K_{(\alpha\beta\gamma)}$ 
in Eq. (\ref{suma3}) \cite{pizdalon}. Hence, we can use Eq. (\ref{masakra}) again and arrive
 at the condition (\ref{2x2x2}).

The price to pay, apart from the mere sufficiency of the condition 
(\ref{2x2x2}), is that the witnesses 
(\ref{skurwisyn}) make no distinction between biseparable 
tripartite reductions 
(which now are not forbidden by the symmetry) and genuine 
3-qubit entangled ones, 
and hence, the inequality (\ref{2x2x2}) indicates only 
general 3-qubit entanglement. However, note that the set of all
biseparable states is closed, and hence each genuine
3-qubit entangled state possesses an open neighborhood
consisting of only genuine 3-qubit entangled states. 
Thus, the criterion (\ref{2x2x2}) also detects genuine 3-qubit entangled 
states in some open vicinity of symmetric states, 
but the size of this vicinity is a priori not known (the same
remark applies to the criterion (\ref{2x2'}) as well).
We will partially solve this drawback using another witnesses
in Section VI. 

\section{An example - Dicke states}

There exists a famous and experimentally accessible family of symmetric states, called Dicke states \cite{Dicke}:
\begin{equation}\label{Psi}
|\Psi_{N,k}\rangle={N \choose k}^{-\jd}\big(|\underbrace{11\dots 1}_{k}000\dots 0\rangle+\text{perm}\big)\,,
\end{equation}
('perm' stands for all possible remaining permutations), which are generalizations
of $N$-qubit $W$-states $|W_N\rangle$:
\begin{equation}
|W_N\rangle=|\Psi_{N,1}\rangle .\label{dableju}
\end{equation}
Dicke states correspond (in spin $1/2$ language) to a fully symmetric flip of $k$ out of $N$ spins. Such states appear in many physical processes, such as superradiance, superfluorescence. As we already mentioned, they can also be realized with photons (where the qubits are encoded in the polarization degree of freedom), or trapped ion systems (where the qubits correspond to two internal  states of ions). In this Section we specify and apply our general results of the previous Sections to the Dicke states.  
We explicitly construct 
for $|\Psi_{N,k}\rangle$ the inequalities (\ref{2x2min}) and (\ref{2x2x2}). In particular, 
 we  derive all 
necessary expressions for the analysis of the experimental data on $7$- and 
$8$-qubit $W$-states, which we will perform in the next Section.   

For practical reasons, we choose the number of excited qubits $k$ to be smaller
 than the integer part of $N/2$. Also note that alternatively the states 
(\ref{Psi}) can be defined as the eigenstates of the total angular 
momentum: 
\begin{equation}\label{total}
|\Psi_{N,k}\rangle =|N/2,N/2-k\rangle.
\end{equation}

We first consider 2-qubit entanglement. The reduced 2-qubit density matrices all 
have the following form:
\begin{equation}
\varrho_2={N\choose k}^{-1}\!\!\!\Big(c_0 |00\rangle\langle 00| + c_1|11\rangle\langle 11| 
+ 2c_+|\phi_+\rangle\langle\phi_+| \Big)\,,
\end{equation}
where
\begin{equation}
|\phi_+\rangle=\frac{1}{\sqrt{2}}(|01\rangle+|10\rangle)\, ,
\end{equation}
and the coefficients are given by the following binomials:
\begin{equation}\label{c2}
c_0={N-2 \choose k},\quad c_1={N-2 \choose k-2},\quad c_+={N-2 \choose k-1}\,.
\end{equation}
In the basis $|00\rangle,|11\rangle,|01\rangle,|10\rangle$ the partially 
transposed matrix $\varrho_2^{T_1}$ is given by:
\begin{equation}
\varrho_2^{T_1}= {N\choose k}^{-1}\!\!\left(\begin{array}{cccc} c_0 & c_+ & 0 & 0 \\
                                                              c_+ & c_1 & 0 & 0 \\
                                                              0 & 0 & c_+ & 0 \\
                                                              0 & 0 & 0 & c_+ \end{array}\right)\,.
\end{equation}
In the generic case, when all the constants from Eqs. (\ref{c2}) are non-zero, 
$\varrho_2^{T_1}$ has one negative eigenvalue:
\begin{equation}
\lambda_-=\frac{1}{2}{N\choose k}^{-1}\big(c_0+c_1-\sqrt{(c_0-c_1)^2+4c_+^2}\big),
\end{equation}
as $c_0c_1-c_+^2<0$, and hence the states (\ref{Psi}) possess bipartite 
entanglement. The normalized 
eigenvector corresponding to $\lambda_-$ is given by:
\begin{eqnarray}\label{vec1}
& & |\psi\rangle=\frac{1}{\sqrt{1+t^2}}\big(|00\rangle-t|11\rangle\big),\label{wektor2q}\\
& & t=\frac{c_0-c_1}{2c_+}+\sqrt{\Big(\frac{c_0-c_1}{2c_+}\Big)^2+1}\,.
\end{eqnarray}
We see that $|\psi\rangle$ is already in the Schmidt decomposed form w.r.t. the chosen basis, and 
hence no unitary rotation $U$ is needed. As that rotation was the only ingredient needed to construct 
the spin squeezing inequalities (\ref{2x2min}) and (\ref{2x2'}) (because the 
angle $\varphi$ is minimized over), we simply put ${\bf k,l,n}=x,y,z$ in them.


Although in theory both inequalities (\ref{2x2min}) and (\ref{2x2'}) are 
equivalent, and we could use the latter due to simplicity, the inequality 
to be measured is rather (\ref{2x2min}) as in real-life experiments one 
does not obtain perfectly symmetric states. Using Eq. (\ref{total}) we find 
that for the perfect Dicke states:
\begin{eqnarray}
& &\langle (J^z)^2\rangle +\frac{N(N-2)}{4} = \frac{N(N-1)}{2}-Nk+k^2, 
\label{2qlhs}
\\
& & \sqrt{\Big[\langle
(J^x)^2\rangle+\langle (J^y)^2\rangle
-\frac{N}{2}\Big]^2+(N-1)^2\langle J^z
\rangle^2} =\nonumber\\
& &\sqrt{(Nk-k^2)^2+\frac{(N-1)^2(N-2k)^2}{4}} \label{2qrhs}.
\end{eqnarray}  
For the experimentally interesting examples of the $7$- and $8$-qubit $W$-states $|W_7\rangle$, $|W_8\rangle$, the expressions (\ref{2qlhs}) 
and (\ref{2qrhs}) take the following values: $15.000$ and $16.155$ respectively for $|W_7\rangle$; $21.000$ and $22.136$ respectively for $|W_8\rangle$. 

Let us now proceed with the analysis of tripartite entanglement. All tripartite 
reductions are of the form:
\begin{eqnarray}
\varrho_3 &=& {N\choose k}^{-1}\!\!\!\Big(\kappa_0 |000\rangle\langle 000| + \kappa_1 |111\rangle\langle 111|+\nonumber\\
& & + 3\omega|W_3\rangle\langle W_3|+3\omega'|W'_3\rangle\langle W'_3| \Big)\,,\label{pizda}
\end{eqnarray}
where $|W'_3\rangle=1/\sqrt{3}\,(|011\rangle+|101\rangle+|110\rangle)$, and
\begin{eqnarray}
& &\kappa_0={N-3 \choose k},\quad \kappa_1={N-3 \choose k-3},\label{par1}\\
& &\omega={N-3 \choose k-1},\quad \omega'={N-3 \choose k-2} \,.\label{par2}
\end{eqnarray}
In the basis $|000\rangle,|110\rangle,|101\rangle,|010\rangle$,$|001\rangle,
|111\rangle$,$|100\rangle,|011\rangle$ the partially transposed matrix $\varrho_3^{T_1}$ reads:
\begin{equation}
\varrho_3^{T_1}= {N\choose k}^{-1}\!\!\left(\begin{array}{cccccccc} \kappa_0 & \omega & \omega & 0 & 0 & 0 & 0 & 0 \\
                                                                   \omega & \omega ' & \omega ' & 0 & 0 & 0 & 0 & 0 \\
                                                                   \omega & \omega ' & \omega ' & 0 & 0 & 0 & 0 & 0 \\
                                                                    0 & 0 & 0 &  \omega & \omega & \omega ' & 0 & 0 \\
                                                                    0 & 0 & 0 &  \omega & \omega & \omega ' & 0 & 0 \\
                                                                    0 & 0 & 0 &  \omega ' & \omega ' & \kappa_1 & 0 & 0\\
                                                                    0 & 0 & 0 & 0 & 0 & 0 & \omega & 0\\
                                                                    0 & 0 & 0 & 0 & 0 & 0 & 0 & \omega'
                                                         \end{array}\right)\,.
\end{equation}
In the generic case it has two negative eigenvalues:
{\setlength \arraycolsep{1pt}
\begin{eqnarray}\label{ne3}
\mu_-&=&\frac{1}{2}{N\choose k}^{-1}\!\!\!\big(\kappa_0+2\omega'-\sqrt{(\kappa_0-2\omega')^2+8\omega^2}\big),\\
\mu'_-&=&\frac{1}{2}{N\choose k}^{-1}\!\!\!\big(\kappa_1+2\omega-\sqrt{(\kappa_1-2\omega)^2+8\omega'^2}\big)\label{ne3'}
\end{eqnarray}}
(because $\kappa_0\omega'<\omega^2$ and $\kappa_1\omega<\omega'^2$), and thus the states 
$|\Psi_{N,k}\rangle$ possess tripartite entanglement as well. Since there are two generically different negative eigenvalues, there will be two different spin squeezing inequalities (\ref{2x2x2}). As before, we will generate them from the eigenvectors corresponding to $\mu_-$ and $\mu'_-$, which read:
\begin{eqnarray}
|\psi\rangle &=& |000\rangle-\alpha|1\rangle \otimes\big(|01\rangle+|10\rangle\big),\label{v1}\\
|\psi'\rangle &=& |111\rangle-\alpha'|0\rangle \otimes\big(|01\rangle+|10\rangle\big),\label{v2}
\end{eqnarray}
where:
\begin{eqnarray}
\alpha &=& \frac{\kappa_0-2\omega'}{4\omega}+\sqrt{\Big(\frac{\kappa_0-2\omega'}{4\omega}\Big)^2+\frac{1}{2}},\\
\alpha' &=& \frac{\kappa_1-2\omega}{4\omega'}+\sqrt{\Big(\frac{\kappa_1-2\omega}{4\omega'}\Big)^2+\frac{1}{2}}.
\end{eqnarray}
The vectors (\ref{v1}), (\ref{v2}) are not normalized, as the norm is irrelevant for the PPT condition (\ref{poly2}).
 After proper rescaling, $|\psi\rangle$ and $|\psi'\rangle$ can be rewritten in the desired form (\ref{W}):
\begin{eqnarray}
& & |\psi\rangle=A\otimes {\bf 1}\otimes {\bf 1} |W_3\rangle , \label{psi}\\
& & |\psi'\rangle=A'\otimes \sigma^x \otimes \sigma^x |W_3\rangle ,  \label{psi'}
\end{eqnarray}
where $A,A' \in SL(2,\mathbb{C})$ are defined as follows:
\begin{eqnarray}
& & A= \pm \left( \begin{array}{cc} 0 & \frac{1}{\sqrt{\alpha}}\\\label{A}
                            -\sqrt{\alpha} & 0 \end{array},\right)\\
& & A'= \pm \left( \begin{array}{cc} \text{i}\sqrt{\alpha'} & 0\\\label{A'}
                             0 & -\frac{\text{i}}{\sqrt{\alpha'}}\end{array}\right).
\end{eqnarray}

Before we proceed with the construction of the inequalities (\ref{2x2x2}), let us note that having the {\it explicit} forms of the negative eigenvalues and the corresponding eigenvectors of $\varrho_3^{T_1}$ it is straightforward to calculate the sum over all triples of qubits (\ref{poly2}). It is just given by:
\begin{eqnarray}
&\sum_{\langle abc \rangle}&\text{tr}_{abc}\big(\varrho_{abc}|\psi\rangle\langle\psi|^{T_1}\big)={N \choose 3}\mu_-||\psi||^2\nonumber\\
&=& {N \choose 3}\mu_-\frac{2\alpha^2+1}{3\alpha}\label{prosta},
\end{eqnarray}
for $\mu_-$ and $|\psi\rangle$, and by the analogous expression for $\mu'_-$ and $|\psi'\rangle$. However, our goal here is to express Eq. (\ref{prosta}) using total angular momentum, in order to make it experimentally available and connect it with the spin squeezing.

Hence, following the procedure described in Section III, we first have to find the Lorentz transformations and rotations generated by matrices from Eqs. (\ref{psi}) and (\ref{psi'}). These transformations are the following: 
matrix (\ref{A}) generates, according to Eq. (\ref{Lorentz}),  
the rotation by $\pi$ around $y$-axis, followed by a boost along $z$-axis:
\begin{eqnarray}
\Lambda(A)&=& \left( \begin{array}{cccc}
\gamma & 0 & 0 & -\gamma\beta\\
0 & -1 & 0 & 0\\
0 & 0 & 1 & 0\\
\gamma\beta & 0 & 0 & -\gamma \end{array}\right),\label{LA}\\
\beta &=& \frac{\alpha^2-1}{\alpha^2+1}, \, \gamma=\frac{1}{\sqrt{1-\beta^2}}\,. \label{L}
\end{eqnarray}
Obviously the identity operator ${\bf 1}$ from Eq. (\ref{psi}) generates the trivial rotation, so we have in this case $R={\bf 1}$. Matrix (\ref{A'}) generates the rotation by $\pi$ around $z$-axis, followed by a boost along it:
\begin{eqnarray}
\Lambda(A') &=& \left( \begin{array}{cccc}
\gamma' & 0 & 0 & \gamma'\beta'\\
0 & -1 & 0 & 0\\
0 & 0 & -1 & 0\\
\gamma'\beta' & 0 & 0 & \gamma' \end{array}\right),\label{LA'}\\
\beta' &=& \frac{\alpha'^2-1}{\alpha'^2+1}, \, \gamma'=\frac{1}{\sqrt{1-\beta'^2}}\,, \label{L'}
\end{eqnarray}
while $\sigma^x$ from Eq. (\ref{psi'}) generates the rotations by $\pi$ around $x$-axis:
\begin{equation}
R(\sigma^x)= \left( \begin{array}{cccc}
1 & 0 & 0 & 0\\
0 & 1 & 0 & 0\\
0 & 0 & -1 & 0\\
0 & 0 & 0 & -1 \end{array}\right).\label{R}
\end{equation}
in the spaces of the second and the third qubit.

\begin{figure}
\includegraphics[height=0.26\textwidth, width=0.238\textwidth]{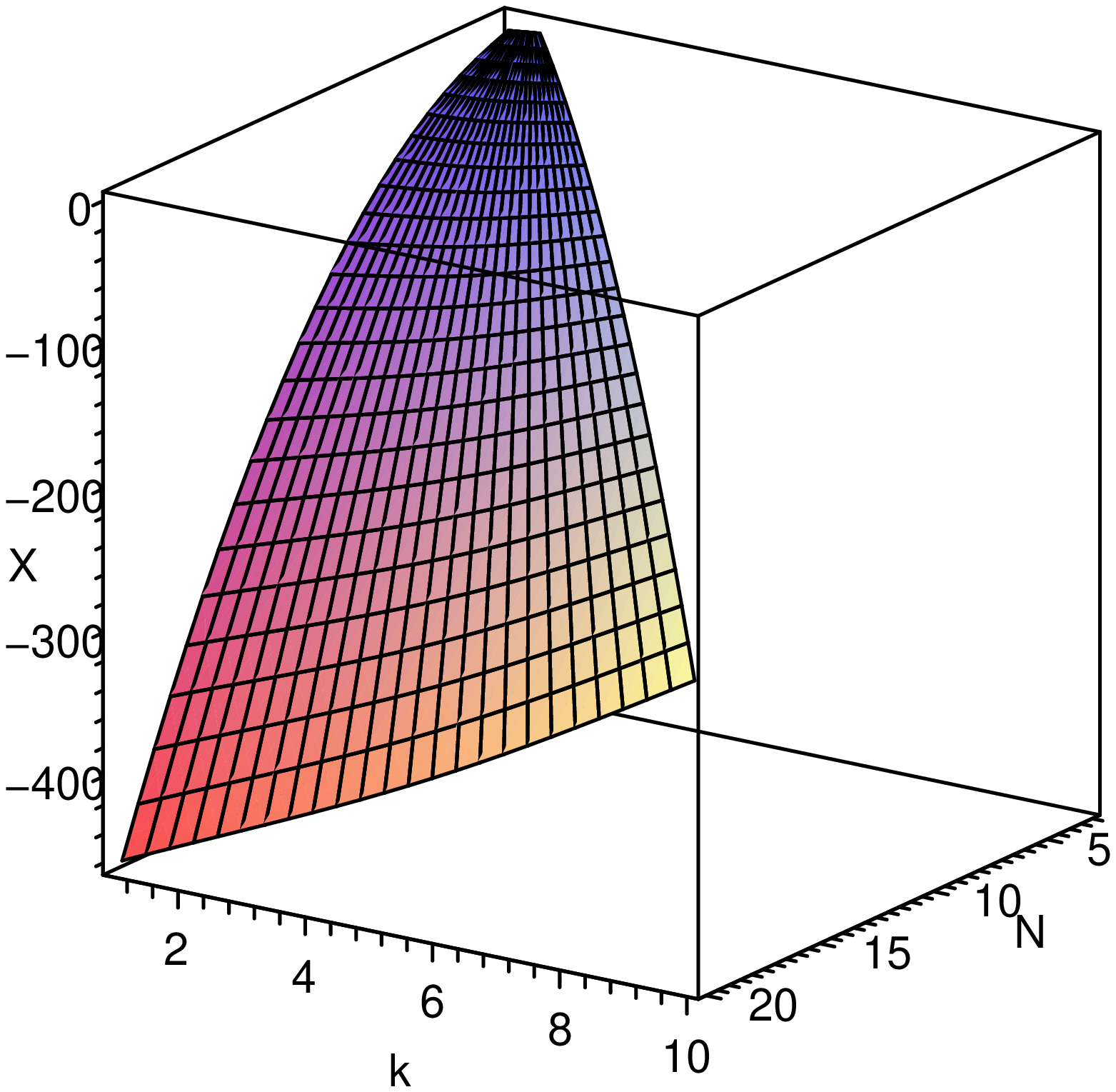}
\includegraphics[height=0.26\textwidth, width=0.238\textwidth]{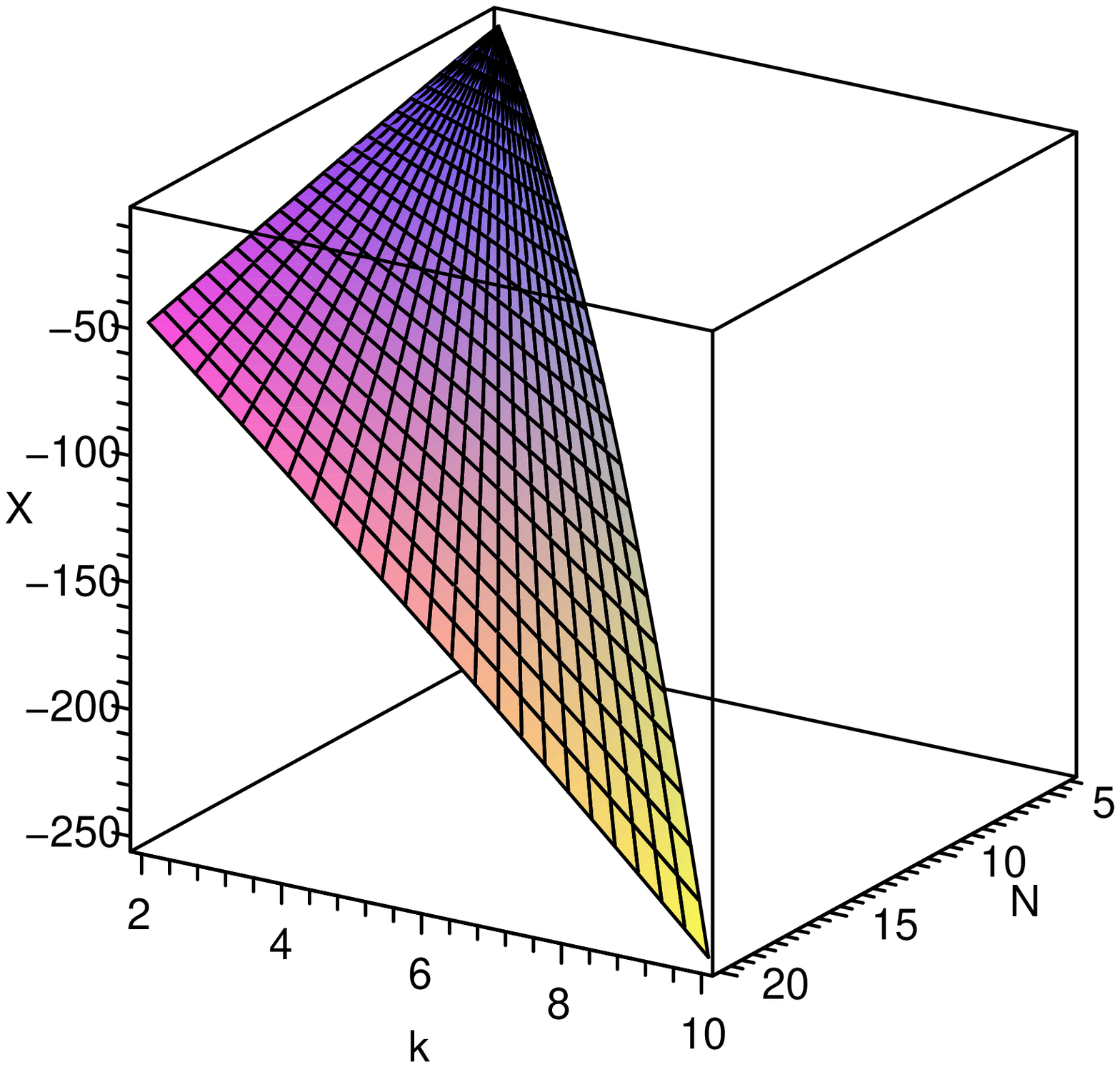}
\caption{(Color online) The (interpolated) plots of the l.h.s. of the inequality (\ref{2x2x2}) corresponding to the eigenvectors $|\psi\rangle$ (left) and  $|\psi'\rangle$ (right).}
\label{figa}
\end{figure}

Next, from the matrices $\Lambda(A)$, $R={\bf 1}$, and $\Lambda(A')$, $R(\sigma^x)$, we construct two copies of the tensor $K_{\alpha\beta\gamma}$, according to Eq. (\ref{KW}). Finally, having $K_{\alpha\beta\gamma}$, we construct the corresponding parameters $X(\Psi_{N,k})$, defined in Eq. (\ref{2x2x2}), and check the 3-qubit spin squeezing inequalities. The resulting expressions are lengthy but straightforward, and hence we will omit them here. Let us stress that for the ideal, generic Dicke states we obtain two independent inequalities, and both of them must be satisfied. Fig.\ \ref{figa} shows the plots of $X(\Psi_{N,k})$ as a function of $N$ and $k$ \cite{12}.

Let us now analyze the $N$-qubit $W$-states $|W_N\rangle$ of Eq. (\ref{dableju}). 
In this case, from Eqs. (\ref{par1}) and (\ref{par2}) we see that $\kappa_0=N-3$, $\omega=1$, and $\kappa_1=\omega'=0$. Substituting this constants into Eqs. (\ref{ne3}) and (\ref{ne3'}), we obtain that there remains only one
 negative eigenvalue of $\varrho_3^{T_1}$ given by $\mu_-$. As a consequence, states $|W_N\rangle$ lead to
 only one spin squeezing inequality, generated by the matrix $\Lambda(A)$ from Eq. (\ref{LA}) and the trivial rotation $R={\bf 1}$. The parameter $\alpha$ from Eq. (\ref{L}) is now equal to:
\begin{equation}
\alpha=\frac{N-3}{4}+\sqrt{\Big(\frac{N-3}{4}\Big)^2+\frac{1}{2}}.
\end{equation}\label{alpha}

\begin{figure}[t]
\includegraphics[height=0.35\textwidth, width=0.48\textwidth]{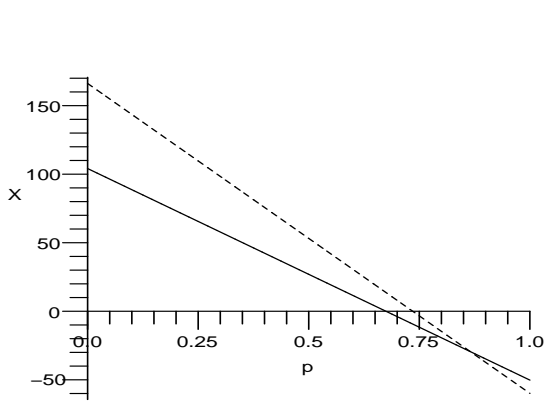}
\caption{The plot of the parameter $X$ as a function of the amount of noise for noisy $W$-states of $N=7$ qubits (solid line) and $N=8$ qubits (dashed line).}
\label{noisefig}
\end{figure}

For the state $|W_7\rangle$ 
we obtain from the corresponding formulas that: 
\begin{equation}\label{expl1}
\Lambda(A)= \left( \begin{array}{cccc}
1.337 & 0 & 0 & -0.888\\
0 & -1 & 0 & 0\\
0 & 0 & 1 & 0\\
0.888 & 0 & 0 & -1.337 \end{array}\right)
\end{equation} 
and $X(W_7)= -44.04$. 

For the state $|W_8\rangle$,
the corresponding matrix is given by: 
\begin{equation}\label{expl2}
\Lambda(A)= \left( \begin{array}{cccc}
1.529 & 0 & 0 & -1.157\\
0 & -1 & 0 & 0\\
0 & 0 & 1 & 0\\
1.157 & 0 & 0 & -1.529 \end{array}\right).
\end{equation}
and the parameter $X(W_8)=\ -59.88$.

To better understand the meaning of the above values of the parameter 
$X(\Psi_{N,k})$, let us briefly consider a less idealized situation and mix 
the states (\ref{Psi}) with the white noise: 
\begin{equation}\label{noise}
\varrho=p|\Psi_{N,k}\rangle\langle \Psi_{N,k}|+(1-p)\frac{{\bf 1}}{2^N}.
\end{equation}
We then calculate the parameter $X(\varrho)$ as if the state (\ref{noise}) were an 
experimental output: we calculate the averages of the spin operators in Eq. (\ref{2x2x2}) 
using the density matrix (\ref{noise}), while plugging the tensor $K_{\alpha\beta\gamma}$ calculated for the ideal Dicke states. Thus, $X(\varrho)=p\,X(\Psi_{N,k})+(1-p)\,X({\bf 1}/2^N)$. 
The results for the states $|W_7\rangle$ and $|W_8\rangle$ are presented in Fig.\ \ref{noisefig}.

\section{Experimental results}

The aim of this Section is to  apply the tools developed in the previous Section to the
recent experiment of Ref. \cite{blatt8}. In this experiment, $7$- and $8$-qubit 
$W$-states have been produced in an ion trap, 
dedicated to quantum information processing
\cite{QC-HowTo}. We begin this Sections by presenting necessary details of the experiment, and follow by applications
of our generalized squeezing inequalities. 

\subsection{Description of the experiment}

\begin{figure}[t]
\begin{center}
\includegraphics[width=0.45\textwidth]{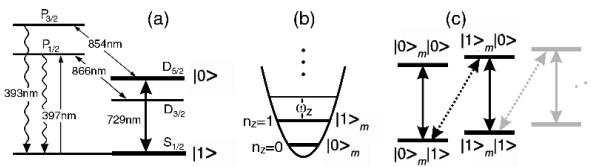}
\caption{\label{levelscheme} (a) Level scheme of Ca$^+$. (b) Schematics of the 
two lowest levels of the harmonic oscillator describing the bus mode. 
(c) Joint energy level diagram of the electronic qubit levels 
\{$\ket{1}$,$\ket{0}$\} and the phonon numbers of the ion's motional mode used for entanglement generation 
\{$\ket{0}_m$,$\ket{1}_m$\}. Carrier transitions
are marked as solid arrows, the blue sideband transition as 
a dashed arrow. Note that the $\ket{0}_m\ket{0}$--level does not 
couple to the blue sideband.}
\end{center}
\end{figure}

Strings of up to eight $^{40}$Ca$^+$ ions are held in a linear ion trap 
capable of storing the ions for several days, a time sufficiently long 
for creating an entangled state more than $10^6$ times. The qubits are 
encoded in superpositions of the S$_{1/2}$  ground state and the 
metastable D$_{5/2}$ state of the Ca$^+$ ions (lifetime of the $D_{5/2}$ 
level: $\tau \approx 1.16$~s). For the atomic level scheme, we refer 
to Fig.~\ref{levelscheme}a. Each ion in the linear string is individually 
addressed by a series of tightly focused laser pulses on the $\ket{1} \equiv
S_{1/2} (m_j=-1/2) \longleftrightarrow \ket{0} \equiv D_{5/2}
(m_j=-1/2)$ quadrupole transition with narrowband laser radiation
near 729~nm. Depending on its frequency, the laser couples either
the states $|n\rangle_m|1\rangle\leftrightarrow
|n\rangle_m|0\rangle$ (carrier pulse) or the states
$|n\rangle_m|1\rangle\leftrightarrow |n+1\rangle_m|0\rangle$ (blue
sideband pulse, laser detuned by $+\omega_z$ w.r.t. the atomic transition, see Fig.~\ref{levelscheme}c). 
Here, $n$ denotes the vibrational quantum number of the ion string's 
center--of--mass motion. Via sideband cooling and optical pumping, the ions 
are prepared in the $\ket{0}_m\ket{11\cdots 1}$--state.

The $N$-ion $W$-states: 
\begin{eqnarray}
|0\rangle_m|W_N\rangle &=& \frac{1}{\sqrt{N}}\sum_i
|\chi_i\rangle, 
\nonumber
\\ 
|\chi_i\rangle &=&|0\rangle_m|x_N\ldots
x_1\rangle, 
\nonumber
\\
x_k&=&\left\{\begin{array}{c}1,\,\mbox{if }
k=i\\0,\, \mbox{if }k\neq i\end{array}\right.\label{wex}
\end{eqnarray}
(note the reverse ordering of the qubits) are created by
applying the sequence of laser pulses shown in Table \ref{Creation_sequence} 
to the ions. 
\begin{table*}
\caption{\label{Creation_sequence}  Creation of a $\ket{W_N}$--state
($N=\{6,7,8\}$). The numbers within the state vector refer to the
phonon excitations of the center--of--mass mode of the ion
crystal. The electronic states are labeled by $|1\rangle$ and
$|0\rangle$. $R^c_n(\theta)$ denotes a carrier pulse of length
$\theta$ applied to the ion $n$, $R^+_n(\theta)$ a blue sideband
pulse. (i1) $\cdots$ (i3) mark initialization steps, (1) $\cdots$
(N) the actual entangling steps. Note that we count the atoms from
right to left.\vspace{0.5cm} }{
\begin{tabular}{c|l}
~ & $\ket{0}_m\ket{111\cdots 1}$ 
\\
(i1) & ${\xrightarrow{R^C_N(\pi)R^C_{N-1}(\pi) \cdots   R^C_1(\pi)}}$
\\
~ & $\ket{0}_m\ket{000\cdots 0}$ 
\\
~& Check state via fluorescence
\\
 (i2) & ${\xrightarrow{R^+_1(\pi)}}$
\\
~ & $\ket{0}_m\ket{000\cdots 0}$ 
\\
~& Check state via fluorescence
\\
 (i3) &${\xrightarrow{R^C_N(\pi)}}$ 
\\
~ & $\frac{1}{\sqrt{N}}\ket{0}_m\ket{100\cdots 0}$ 
\\
 (1) &${\xrightarrow{R^+_N(2\arccos(1/\sqrt{N})}}$ 
\\
~ & $\frac{1}{\sqrt{N}}\ket{0}_m\ket{100\cdots 0} + 
\frac{\sqrt{N-1}}{\sqrt{N}}\ket{1}_m\ket{000\cdots 0}$
\\
 (2) &${\xrightarrow{R^+_{N-1}(2\arcsin(1/\sqrt{N-1})}}$ 
\\
~ & $\frac{1}{\sqrt{N}}\ket{0}_m\ket{100\cdots 0} +
\frac{1}{\sqrt{N}}\ket{0}_m\ket{010\cdots 0}+\frac{\sqrt{N-2}}
{\sqrt{N}}\ket{1}_m\ket{000\cdots 0}$ 
\\
\vdots & ~ ~ ~ ~ ~\vdots
\\
~ & $\frac{1}{\sqrt{N}}\ket{0}_m\ket{100\cdots 0} +
\frac{1}{\sqrt{N}}\ket{0}_m\ket{010\cdots 0}+
\cdots+\frac{1}{\sqrt{N}}\ket{1}_m\ket{000\cdots 0}$
\\
 (N) &${\xrightarrow{R^+_{1}(2\arcsin(1/\sqrt{1})}}$ 
\\
~ & $\frac{1}{\sqrt{N}}\ket{0}_m\ket{100\cdots 0} 
+\frac{1}{\sqrt{N}}\ket{0}_m\ket{010\cdots 0}+
\cdots+\frac{1}{\sqrt{N}}\ket{0}_m\ket{000\cdots 1}$
\\
\end{tabular}}
\bigskip\bigskip
\end{table*}
First, the $\ket{0}_m\ket{111\cdots 1}$--state is prepared by $N$
$\pi$--pulses on the carrier transition applied to ions \#1 to
\#$N$. Then, laser light coupling the $|1\rangle$ state resonantly
to the short-lived excited state $P_{1/2}$ projects the ion string
on the measurement basis. Absence of fluorescence reveals whether
all ions were prepared in $\ket{0}$. Similarly, we test the
motional state with a single blue $\pi$ pulse. Absence of
fluorescence during a subsequent detection period indicates ground
state occupation. This initialization procedure can be viewed as a
generalized optical pumping with the target state
$\ket{0}_m\ket{11\cdots 1}$. If both checks were successful (total
success rate $\geq 0.7$), we continue with the
$\ket{W}$--preparation at step ($i3$) in
Tab.~\ref{Creation_sequence} to create the state
$\ket{0}_m\ket{10\cdots 0}$. The entangling procedure starts 
by moving most of the population to the
$\ket{1}_m\ket{000\cdots 0}$ with a blue sideband pulse of pulse area
$\theta_N=2\arccos(1/\sqrt{N}) $
leaving $1/N$ of the population back in $\ket{0}_m\ket{100\cdots 0}$.
Now, $W$--states are efficiently generated by redistributing the
$\ket{1}_m\ket{0\ldots 0}$ state population equally among the states
$\ket{0}_m\ket{0\ldots 01_i0\ldots 0},i=1\ldots N-1$. This is
achieved by $N-1$ blue sideband pulses of pulse length
$\theta_i=2\arcsin(1/\sqrt{N-i})$.
Note that for an ion string in the motional ground state, 
blue--sideband
pulses acting on an ion in the $\ket{0}$--state have no effect. 
We note that this production
method scales quite advantageous, as the required sideband pulse
area increases only logarithmically. Therefore, even large
$W$--states can be created quite efficiently.

Furthermore, the space spanned by the states $|\chi_i\rangle $ 
from Eq. (\ref{wex}) is
decoherence-free w.r.t. the collective dephasing, which is the
main decoherence mechanism in the setup \cite{Haeffner05A}.
Therefore, the $W$--states are quite robust and long-lived. In
addition, even during the creation of a $W$--state, the energy
difference between all superpositions never exceeds more than the
one of a single qubit. Thus, the requirements to laser frequency
and magnetic field noise for a high--fidelity generation of
$W$--states are rather modest. We discuss the imperfections
in a separate section below.

\subsection{State tomography}
\label{FidReconstr}
Information about the $N$-ion quantum state is obtained by exciting
the ion string on the $S_{1/2}\leftrightarrow P_{1/2}$ transition
(see Fig.~\ref{levelscheme}a),
and detecting the ion's fluorescence spatially resolved with a CCD
camera state \cite{QC-HowTo}. The measurement of an ion's
fluorescence amounts to measuring the Pauli matrix $\sigma^z$,
if $|0\rangle$ and $|1\rangle$ are identified with the eigenstates
of $\sigma^z$. The measurement of $\sigma^x (\sigma^y)$ is
accomplished by applying a suitable $\pi/2$ carrier pulse to the
ion prior to the state detection \cite{roos04}.

\begin{figure*}
\begin{center}
\includegraphics[width=0.6\textwidth,angle=-90]{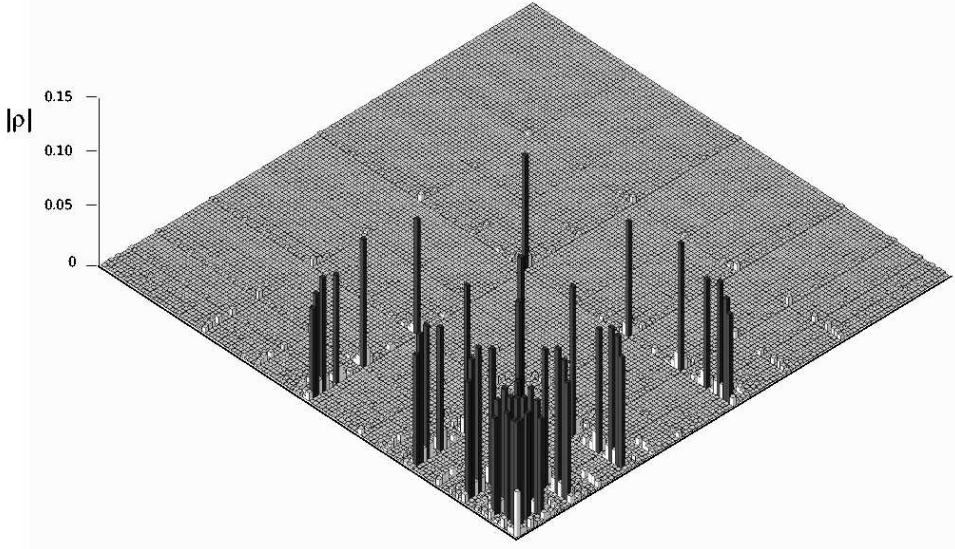}
\caption{Absolute values of the reconstructed density matrix of a
$\ket{W_7}$--state as obtained from quantum  state tomography.
Ideally, the dark entries should all have the same height of
$\frac{1}{7}$, the bright bars should vanish.} \label{tomo7W}
\end{center}
\end{figure*}

To verify the entanglement of the produced state, a
measurement of a witness operator, yielding a negative expectation
value, would be sufficient in principle. However, the optimal witness is a
priori not known. Therefore, it can be advantageous to get as much
information as possible about the produced quantum state. Full
information on the $N$--ion entangled state is obtained via
quantum state reconstruction. For this we expand the density
matrix in a basis of observables, and measure the corresponding
expectation values. For the basis, we choose tensor products of
Pauli matrices:
$\sigma^{i_N}_{N}\otimes\ldots\otimes\sigma^{i_1}_{1}$ 
(note the reverse ordering). We use
$3^N$ different bases and repeat the experiment 100 times for each
basis. For $N=8$, we need thus $656\,100$ experiments and a total
measurement time of 10 hours. We follow the iterative procedure
outlined in Ref. \cite{Hradil04} for performing a
maximum--likelihood estimation of $\rho.$ Other reconstruction 
methods would also be possible \cite{braunschweig}. The procedure 
ensures also positivity of the reconstructed matrix. The resulting 
matrix for the state $|W_7\rangle$  is displayed in Fig.~\ref{tomo7W},
the numerical values are published in the on-line material of 
Ref.~\cite{blatt8}.

A Monte Carlo simulation is used to estimate uncertainties in the
density matrix elements, and in quantities derived from it, that are
due to quantum noise in the state reconstruction measurements:
starting from the reconstructed density matrix, we simulate the
measurement process and reconstruct up to 100 times the density
matrix from these simulated measurements. From the set of
reconstructed density matrices, the spread in the expectation
values of the observable of interest can be estimated. For density
matrices close to pure states, we observe that the purity of the
reconstructed matrices often slightly decreases (for the
$|W\rangle$ states by about 2\%). Therefore, we conclude that the
reconstruction process rather underestimates the entanglement in
the experimentally produced quantum states.



\subsection{Experimental imperfections}
For an investigation of the experimental imperfections and
scalability, we simulate the preparation procedure by solving the
Schr\"odinger equation with the relevant imperfections.

The four major sources of deviations from the ideal $W$--states are
addressing errors, imperfect optical pumping, non--resonant
excitations, and phase noise (laser frequency and magnetic field
noise). For the large $W$--states, we approximate the ions as
two-level systems and include only the first three levels of the
center-of-mass excitation. For a serious analysis of the
imperfections this is by no means sufficient as e.g. no
environment is included. Still the simulation time for the
generation of a $\ket{W_8}$--state under these idealized
conditions is already 20 minutes on a 3 GHz processor using
Matlab. As the computational time for the simulations scales with
$4^N$, it is quite demanding to include a reasonable environment
or even use a density matrix approach.

The fidelity reduction of $\ket{W_6}$ for the different
imperfections are as follows: 0.1 (addressing error), 0.07
(off--resonant excitations), 0.04 [laser frequency noise (200~Hz
rms)]. We note here that as opposed to e.g. experiments on
teleportation \cite{riebe04} or the Cirac--Zoller
controlled--NOT \cite{QC-HowTo}, phase noise (caused by the laser 
frequency noise or magnetic field noise) contributes here much less.
Another possible error source is imperfect ground state cooling.
Intensity noise of the 729--laser ($\Delta I_{\rm max}/I \approx
0.03$) does not contribute significantly. Finally, we
experimentally  observed non--ideal optical pumping which can
result in a reduction of 0.02 of the fidelity per ion. For $N\geq
6$, we therefore minimize the errors due to optical pumping and a
part of the addressing errors by checking the initialization
procedure with a detection sequence (see
Tab.~\ref{Creation_sequence}). Due to this improvements, the
addressing error reduces the fidelity of the $\ket{W_6}$ state by
only 0.05 and the optical pumping errors are basically excluded.
Furthermore, we switched the blue-sideband pulses
adiabatically w.r.t. the trap frequency, such that Fourier components
at the carrier transition do not lead to
off-resonant excitations. Taking this new situation into account,
the fidelity should be of the order of 0.91. Even though it is hard
to estimate the expected
fidelity for $N=8$, it seems that the discrepancy
between the model and the experiment is even larger for $N=8$.
A small part of these discrepancies could be due to the
quantum projection noise in the measurement process as described in
the section on state tomography. However, looking at the density matrices
in detail, we observe that the $\ket{000\cdots 0}$ state seems
always quite strongly populated, especially for large $N$. So far
we have no good explanation for this.

\subsection{Evaluation of the data}

In Ref.~\cite{blatt8} it has already been shown that the states 
are genuine multipartite entangled, multipartite distillable and 
also that all the reduced two-qubit states are entangled. Now we 
want to apply our criteria to the experimental density matrices  
$\varrho_{ex}.$

We begin with the $7$-qubit states. In this case,
the fidelity of the produced states was $F_7=0.763$. To check the
presence of bipartite entanglement, we use the inequality (\ref{2x2min}) 
rather than (\ref{2x2'}), as the experimental states are not symmetric 
due to the experimental imperfections described in the previous subsection.
 According to the theoretical 
analysis of Section IV (c.f. formula (\ref{wektor2q})), 
the frame directions ${\bf k,l,n}=x,y,z$. We find that:
\begin{eqnarray}
& &\langle (J^z)^2\rangle +\frac{N(N-2)}{4} = 14.666\pm 0.016\,,
\nonumber
\\
& & \sqrt{\Big[\langle
(J^x)^2\rangle+\langle (J^y)^2\rangle
-\frac{N}{2}\Big]^2+(N-1)^2\langle J^z
\rangle^2}\nonumber\\ 
& &\quad\quad\quad\quad\quad\quad\quad\quad\quad = 15.148 \pm 0.023\,,
\nonumber
\end{eqnarray}
which clearly proves the presence of bipartite entanglement
in the produced states. 

Let us move to the tripartite entanglement.
We evaluate $X(\varrho_{ex})$ using the Lorentz matrix (\ref{expl1}). 
We find that:
\begin{equation}
X(\varrho_{ex}) = -24.937 \pm 0.202\,,
\label{ex7qb}
\end{equation}
and hence the spin squeezing inequality (\ref{2x2x2}) is fulfilled.  
However, as we mentioned at the end of Section III,
the validity of the inequality (\ref{2x2x2}) only 
proves the presence of {\it some} form of tripartite entanglement
and a priori we do not know if it is genuine 3-qubit entanglement.

Let us now discuss the eight qubit case. Here, the experimentally
reached fidelity was $F_8=0.7215$. The evaluation of the bipartite
criteria yields:
\begin{eqnarray}
& & \langle (J^z)^2\rangle +\frac{N(N-2)}{4} = 20.462 \pm 0.007\,,
\nonumber
\\
& & \sqrt{\Big[\langle
(J^x)^2\rangle+\langle (J^y)^2\rangle
-\frac{N}{2}\Big]^2+(N-1)^2\langle J^z
\rangle^2}\nonumber\\
& &\quad\quad\quad\quad\quad\quad\quad\quad\quad = 20.838 \pm 0.009\,,
\nonumber
\end{eqnarray}
and the tripartite criterion gives:
\begin{equation}
X(\varrho_{ex}) = -29.017 \pm 0.2623.
\label{ex8qb}
\end{equation}
Thus both criteria detect entanglement again.

\section{Simplified criteria for the 3-qubit entanglement}

The general form of squeezing inequalities of Sections II and III is complicated and remains such even when specified to Dicke states. 
It is therefore desirable to derive  alternative inequalities, which are weaker, but have a simple form. 
In fact, in  Ref. \cite{naszprl} it was  proposed to use less general witnesses, developed in Ref. \cite{3qbit}, than those provided by the PPT criterion for the 3-qubit case:
\begin{eqnarray}
& &\mathcal{W}_{GHZ}=\frac{3}{4}{\bf 1}-|GHZ_3\rangle\langle GHZ_3|,\label{wghz}\\
& &\mathcal{W}_{W_1}=\frac{2}{3}{\bf 1}-|W_3\rangle\langle W_3|,\label{ww1}\\
& &\mathcal{W}_{W_2}= \frac{1}{2}{\bf 1}-|GHZ_3\rangle\langle GHZ_3|\label{ww2}\,,
\end{eqnarray}
where now we allow the vectors $|GHZ_3\rangle$ and $|W_3\rangle$ to be defined in an arbitrary 
frame ${\bf k},{\bf l}, {\bf n}$, the same for all three qubits. Apart from the simplicity, 
the advantage of such an approach over the general criterion (\ref{2x2x2}) is that the above 
witnesses detect genuine 3 qubit entanglement in generic states. 

We derived the spin squeezing inequalities corresponding to $\mathcal{W}_{GHZ},\mathcal{W}_{W_1},\mathcal{W}_{W_2}$ using the same technique as in Section III: we expressed the sums 
\begin{equation}\label{sw}
\sum_{\langle abc\rangle}\text{tr}_{abc}\big(\varrho_{abc}\mathcal{W}_{abc}\big)=\text{tr}\big(\varrho\sum_{\langle abc\rangle}\mathcal{W}_{abc}\big) 
\end{equation}
with the total spin operators (\ref{S}). However, instead of using the general formula (\ref{masakra}), we calculated explicitly the occurring products of Pauli matrices (or in other words we used special cases of Eq. (\ref{masakra})). This led us to the following sufficient criteria for the GHZ-type entanglement \cite{corr3}:
\begin{eqnarray}
& & -\frac{1}{3}\langle J_{\bf k}^3\rangle+\langle J_{\bf l}J_{\bf k}J_{\bf l}\rangle-\frac{N-2}{2}\langle J_{\bf n}^2\rangle+\frac{1}{3}\langle J_{\bf k}\rangle\nonumber\\
& & +\frac{N(N-2)(5N-2)}{24}<0\label{ss1}\,,
\end{eqnarray}
and for the GHZ- or W-type entanglement:
\begin{eqnarray}
& & \langle J_{\bf n}^3\rangle-2\langle J_{\bf l}J_{\bf n}J_{\bf
l}\rangle-2\langle J_{\bf k}J_{\bf n}J_{\bf k}\rangle\nonumber\\
& & -\frac{N-2}{2}\big(2\langle J_{\bf k}^2\rangle+2\langle J_{\bf
l}^2\rangle-\langle J_{\bf
n}^2\rangle\big)-\frac{N^2-4N+8}{4}\langle
J_{\bf n}\rangle\nonumber\\
& & +\frac{N(N-2)(13N-4)}{24}<0,\label{ss2}\\
& &-\frac{1}{3}\langle J_{\bf k}^3\rangle+\langle J_{\bf l}J_{\bf k}J_{\bf l}\rangle-\frac{N-2}{2}\langle J_{\bf n}^2\rangle+\frac{1}{3}\langle J_{\bf k}\rangle\nonumber\\
& &+\frac{N^2(N-2)}{8}<0\label{ss3}\,.
\end{eqnarray}

The witnesses (\ref{wghz}-\ref{ww2}) still have a disadvantage that in the sums $\sum_{\langle abc\rangle}\mathcal{W}_{abc}$, the identity gives the dominant contribution, and hence the bigger the system the less sensitive the witnesses become. One possible method to partially overcome this problem is to project the witnesses (\ref{wghz}-\ref{ww2}) onto the symmetric subspace of the space of three qubits:
\begin{eqnarray}
& &\widetilde{\mathcal{W}}_{GHZ}=\frac{3}{4} P_3 -|GHZ_3\rangle\langle GHZ_3|\label{Pwghz},\\
& &\widetilde{\mathcal{W}}_{W_1}=\frac{4}{9} P_3 -|W_3\rangle\langle W_3|,\label{Pww1}\\
& &\widetilde{\mathcal{W}}_{W_2}= \frac{1}{2} P_3 -|GHZ_3\rangle\langle GHZ_3|\label{Pww2}\,,
\end{eqnarray}
where: 
\begin{equation}
P_3=|000\rangle\langle 000| +|111\rangle\langle 111| + |W_3\rangle\langle W_3|+ |W'_3\rangle\langle W'_3| .
\end{equation}
The factor $4/9$ in the definition (\ref{Pww1}) is the maximum overlap between $|W_3\rangle$ and symmetric 
separable states (there are no symmetric biseparable states due to the symmetry) \cite{wei}. The criteria
that such improved witnesses lead to, read respectively: 
\begin{eqnarray}
& & -\frac{1}{3}\langle J_{\bf k}^3\rangle+\langle J_{\bf l}J_{\bf k}J_{\bf l}\rangle+\frac{N-2}{2}\langle J_{\bf k}^2+J_{\bf l}^2\rangle+\frac{1}{3}\langle J_{\bf k}\rangle\nonumber\\
& & +\frac{N(N-2)(N-4)}{12}<0,\label{ss1'}\\
& & \langle J_{\bf n}^3\rangle-2\langle J_{\bf l}J_{\bf n}J_{\bf
l}\rangle-2\langle J_{\bf k}J_{\bf n}J_{\bf k}\rangle\nonumber\\
& & +\frac{N-2}{9}\Big(\frac{25}{2}\langle J_{\bf n}^2\rangle-\langle J_{\bf l}^2+J_{\bf k}^2\rangle\Big)-\frac{N^2-4N+8}{4}\langle
J_{\bf n}\rangle\nonumber\\
& & +\frac{7N(N-2)(N-4)}{72}<0,\label{ss2'}\\
& &-\frac{1}{3}\langle J_{\bf k}^3\rangle+\langle J_{\bf l}J_{\bf k}J_{\bf l}\rangle+\frac{N-2}{12}\big(2\langle J_{\bf k}^2+J_{\bf l}^2\rangle-\langle J_{\bf n}^2\rangle\big)\nonumber\\
& &+\frac{1}{3}\langle J_{\bf k}\rangle+\frac{N(N-2)(N-4)}{48}<0\label{ss3'}\,.
\end{eqnarray}

The potential advantage of using  
$\widetilde{\mathcal{W}}_{GHZ},\widetilde{\mathcal{W}}_{W_1}, \widetilde{\mathcal{W}}_{W_2}$ 
instead of $\mathcal{W}_{GHZ},\mathcal{W}_{W_1},\mathcal{W}_{W_2}$ manifests itself only for
 non-symmetric states. For symmetric states, both families give the same results 
 (apart from $\widetilde{\mathcal{W}}_{W_1}$ due to the factor $4/9$), 
as we can always substitute $\varrho_{abc}$ with $P_3\varrho_{abc} P_3$ in Eq. (\ref{sw}).

Let us apply the above witnesses to the Dicke states $|\Psi_{N,k}\rangle$ of Section IV. As one can easily see from Eq. (\ref{pizda}), only $\mathcal{W}_{W_1}$ and $\widetilde{\mathcal{W}}_{W_1}$ have a chance to detect genuine tripartite entanglement, but not for all $N$ and $k$. For example, for $\ket{W_N}$, $\mathcal{W}_{W_1}$ detects entanglement only for $N\le 4$, and  $\widetilde{\mathcal{W}}_{W_1}$ -- only for $N\le 6$.

\section{Conclusions}
We develop in more detail the novel method 
of detecting entanglement in multiqubit systems, 
first introduced in Ref. \cite{naszprl}, and apply it to the output of
the recent ion trap experiment from Ref. \cite{blatt8}. We also present some novel details regarding that experiment. 
The detection method is based on the use of entanglement witnesses together 
with the concept of spin squeezing. We show in detail 
and on the example, how to
obtain sufficient (or necessary in sufficient for symmetric states)
entanglement criteria for 
detecting 2- and 3-qubit entanglement (or genuine 3-qubit entanglement 
for symmetric states and states sufficiently close to them). We use them to analyze concrete experimental data.
We also provide a novel, mathematically exact
justification for the intuitive picture linking the presence of 
 spin squeezing with the 
non-classical 2-qubit correlations in the system. 
Therefore, our criteria generalize 
the standard notion of spin squeezing as the measure of entanglement in multiqubit systems. 

As the concrete example we study the family of Dicke states and show step-by-step
how our method works in practice. We obtain ready-to-use expressions (\ref{expl1}), (\ref{expl2}) and then apply them to the study of the experimentally generated 
7- and 8-qubit atomic $W$-states from Ref. \cite{blatt8}. In the experimental part, we explain in detail the full state-creation algorithm (Table \ref{Creation_sequence}) and the state tomography procedure used in the experiment from Ref. \cite{blatt8}. We also present the reconstructed $7$-qubit $W$-state in Fig. \ref{tomo7W} and provide the analysis of the experimental imperfections.

Finally, we provide improved sufficient criteria --- Eqs. (\ref{ss1'}-\ref{ss3'}), detecting genuine 3-qubit entanglement, suitable for macroscopic systems.

We thank  A. Ac\'\i n, W. D\"ur, J. Eschner, M. Mitchell, G. T\'oth and especially J. I. Cirac for discussions. We gratefully acknowledge the financial support of: Austrian Science Fund
(FWF), Deutsche Forschungsgemeinschaft
(SFB 407, 436 POL), European Commission (QGATES, CONQUEST PROSECCO,QUPRODIS and OLAQUI networks), ESF PESC QUDEDIS, EU IP Programme ``SCALA'', Institut f\"ur
Quanteninformation GmbH, and MEC (Spanish Goverment) under contract FIS2005-04627. This material is based upon work supported in part
by the U. S. Army Research Office.


\begin{thebibliography}{99}
\bibitem{notka} Note that macroscopic atomic ensembles
may be prepared in symmetric subspace, although
ultimately individual spontaneous emission acts take
the system out of this subspace. Nevertheless, 
especially in mesoscopic systems, the
symmetric component might remain significant for relatively
 long times.
\bibitem{corr1} This is the correct version of the formula (17) from Ref. \cite{naszprl}, 
where we mistakenly omitted few terms.
\bibitem{corr2} The formulas (\ref{masakra}), (\ref{2x2x2}) are the correct 
versions of the formulas (18), (19) of Ref. \cite{naszprl}.
\bibitem{pizdalon} Note that the witnesses (\ref{skurwisyn}) are not equal 
to $P|\psi\rangle\langle\psi|^{T_1}\!P$, which would merely lead to cutting the symmetric part from $\varrho$.
\bibitem{12} The values on the plot in Fig. \ref{figa} differ by the factor 12 w.r.t. Eq. (\ref{prosta}), 
as we have omitted it when going from Eq. (\ref{poly2}) to Eq. (\ref{2x2x2}).
\bibitem{corr3} This is the corrected formula (23) of Ref. \cite{naszprl}.
\end{thebibliography}
\end{document}